\documentclass[twocolumn,aps,prb,groupedaddress,showpacs,altaffilletter,showkeys]{revtex4}

\usepackage{graphicx}

\newcommand{\dif}{\mathrm{d}}
\newcommand{\ij}{\mathrm{j}}

\newcommand\nodata{ ~$\cdots$~ }

\begin{document}

\title{Complex Impedance as a Diagnostic Tool for Characterizing
  Thermal Detectors}

\author{John E. Vaillancourt}
\altaffiliation{Current address: Enrico Fermi Institute, University of
  Chicago, 5640 S. Ellis Ave., Chicago, IL 60637;
  johnv@oddjob.uchicago.edu}
\affiliation{Physics Department, University of Wisconsin, 1150
University Ave., Madison, WI 53706}

\preprint{2005, Rev.\ Sci.\ Instrum., 76, in pressv}

\begin{abstract}
The complex ac impedance of a bolometer or microcalorimeter detector
is easily measured and can be used to determine thermal time
constants, thermal resistances, heat capacities, and sensitivities.
Accurately extracting this information requires an understanding of
the electrical and thermal properties of both the detector and the
measurement system.
We show that this is a practical method for measuring parameters in
detectors with moderately complex thermal systems.
\end{abstract}

\pacs{
  07.20.Fw; 
  07.57.Kp; 
  84.37.+q; 
  85.25.Am;  
  85.30.De 
}
\keywords{bolometers; microcalorimeters; impedance}

\maketitle

\section{Introduction} \label{sect-intro}

Thermal detectors are used in a number of fields ranging from particle
and plasma physics to astrophysics.  The two commonly used forms of
these detectors are bolometers to measure incident power and
microcalorimeters to measure total energy.  Despite these different
applications their construction and operation principles are
similar. In general, they are composed of an absorbing element to
collect incident radiation or particles, a resistive thermometer
coupled to the absorber, and a weak thermal link connecting the
thermometer to a heat sink.

\begin{figure}[t]
\includegraphics[width=9cm]{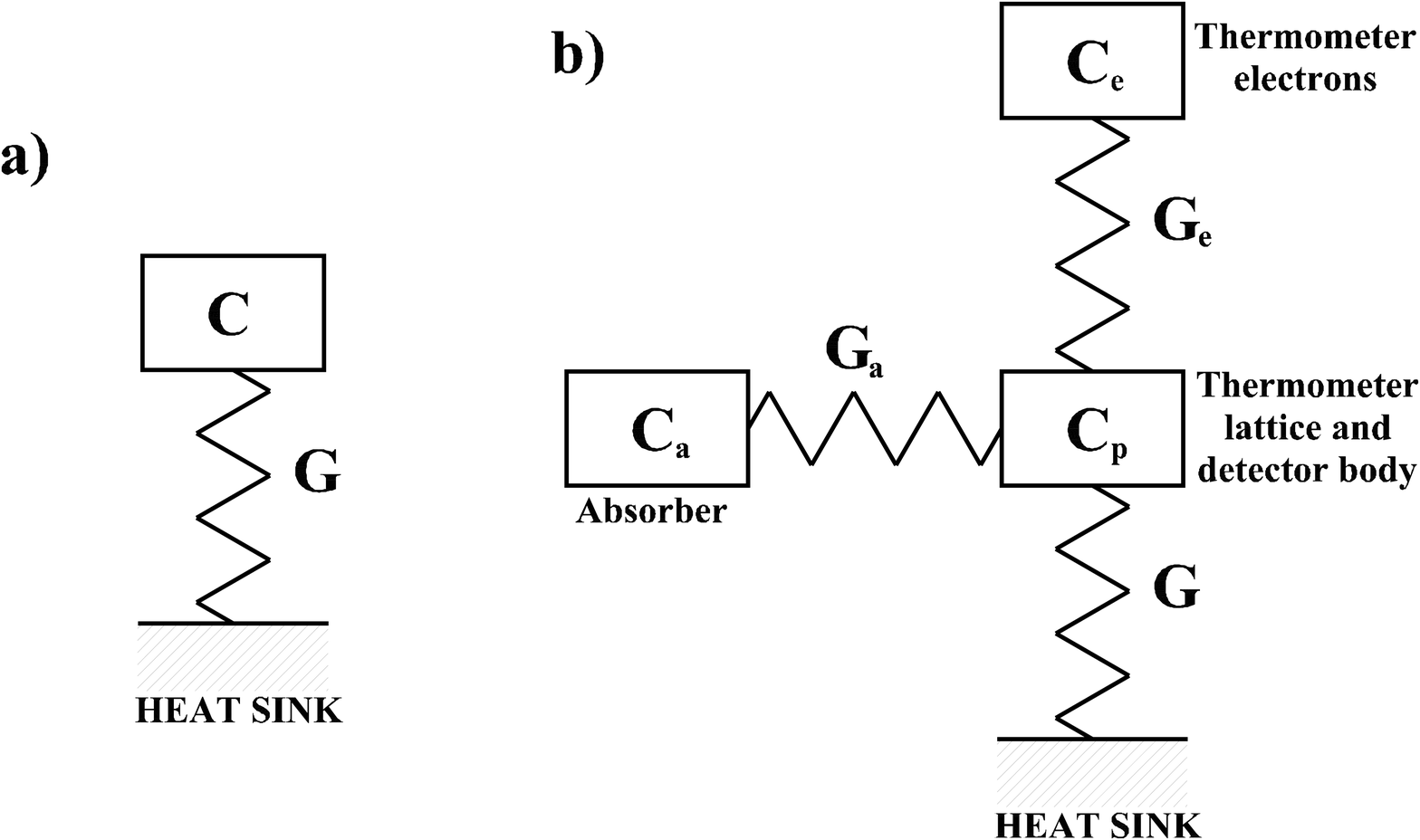}%
\caption{a) The ideal model of a thermal detector consists of a lumped
  heat capacity $C$ connectecd to a heat sink through a weak thermal
  link $G$. b) A more realistic model that includes internal
  couplings.\citep{cal_model}
\label{fig:thermal_models}}
\end{figure}

The simple model treated in most analytical descriptions consists of a
lumped heat capacity connected to a heat sink through a weak thermal
link [Fig.\ \ref{fig:thermal_models}(a)].  The theory predicting the
responsivity, noise properties, and energy resolution of these
detectors has been thoroughly developed.
\citep{jones53,mather82,mather84,mmm84}  However, more complex
thermal models are often needed to describe real detectors. The
internal construction of these detectors often introduces additional
time constants and noise terms which can have major detrimental
effects on performance.\citep{cal_model,biabsorber,masltd10,paul93}

A detector's performance can be predicted and optimized for specific
applications\citep{mather84-2,mmm84,masltd10,bandlerltd10} if the
thermal and electrical properties of their components are known.
Thermal conductivities, heat capacities, and sensitivities can be
extracted from measurements of the resistance vs.\ temperature
relation, dc $IV$ (current-voltage) curves, and time
constants.\citep{jones53,mather84,murari04,giannone02}
The thermal time constant can be found by measuring either the
detector's impulse response function in the time
domain\citep{paul93,schivell82,giannone02,gu98} or its complex ac
impedance $Z(\omega) \equiv \dif V/\dif I$ in the frequency
domain.\citep{mather82,mather84} In this paper we discuss the
impedance method, which can often be used to find more precise
parameter values than the impulse response method.\citep{lindeman04}


We will show that the internal heat capacities and thermal
conductivities of both simple and complex detectors, along with the
thermometer temperature sensitivity and thermometer voltage (or
current) sensitivity can be determined from measurements of dc
$IV$-curves and the ac impedance.  As examples of this technique we
fit impedance data to the ideal detector model and one of the more
complex thermal circuits presented by \citet{cal_model}.  This complex
model includes electron-phonon
decoupling\citep{ltd9dahai,zhang98,hrp9} in the thermometer and a
thermal resistance between the absorber and thermometer.

We begin with a review of the impedance of the simplest detector model
in Sec.\ \ref{sect-simple}\@.  Section \ref{sect-imped_meas} presents
example fits of simple and complex thermal models to impedance data.
In Sec.\ \ref{sect-strays} we discuss the effects of stray electrical
capacitances and inductances in the bias and readout circuits.  While
the examples presented here utilize a voltage readout circuit, the
equations can readily be transformed to current readout.

\section{The Simple Thermal Detector} \label{sect-simple}

The simplest, or \emph{ideal}, detector consists of a single lumped
heat capacity $C$ connected to a heat sink through a weak thermal
link of conductivity $G$ [Fig.\ \ref{fig:thermal_models}(a)].  This
model has been widely discussed by other authors;
\citep{jones53,mather82,mather84,cal_model} we briefly review their
results here.

The dynamic impedance for this simple model is given by
\begin{equation}
Z(\omega) = \frac{R}{(1-\beta_v)}
\frac{1+L_v+\ij\omega\tau}{1-L_v/(1-\beta_v)+\ij\omega\tau}
\label{eq-idealz1}
\end{equation}
where
\begin{eqnarray}
L_v & = & \mathrm{dimensionless~gain} \equiv \frac{\alpha_v P}{GT}
\label{eq-l}\\
R & = & \mathrm{detector~resistance} \nonumber \\
T & = & \mathrm{detector~temperature} \nonumber \\
P & = & \mathrm{Joule~power~dissipated~in~detector}=V^2/R \\
G & = & \mathrm{thermal~conductivity~to~heat~bath} = \dif P/\dif T \\
\tau & = & \mathrm{thermal~time~constant} = C/G \label{eq:tau} \\
\omega & = & \mathrm{angular~frequency} \nonumber\\
\alpha_v & = & \mathrm{thermometer~sensitivity~at~constant~voltage}
\nonumber\\ & = & \left.\frac{T}{R}\frac{\partial R}{\partial T}\right\vert_V\\
\beta_v & = & \mathrm{thermometer~voltage~dependence~at~constant~}T
\nonumber\\ & = & \left.\frac{V}{R}\frac{\partial R}{\partial V}\right\vert_T
\end{eqnarray}
For a linear thermometer the resistance is dependent only on its
temperature and $\beta_v = 0$.  However, this is not generally the
case so we will retain the term here.

Equation (\ref{eq-idealz1}) can be rewritten as
\begin{equation}
Z(\omega) = \frac{Z_0+Z_\infty}{2} + \frac{Z_0-Z_\infty}{2} \frac{1-\ij\omega\bar{\tau}_z}{1+\ij\omega\bar{\tau}_z}
\label{eq-idealz2}
\end{equation}
where
\begin{eqnarray}
Z_\infty & \equiv & \lim_{\omega\rightarrow\infty} Z(\omega) =
R/(1-\beta_v)\mathrm{,} \label{eq-zinf} \\
Z_0 &\equiv & Z(0) = Z_\infty \frac{1+L_v}{1-L_v/(1-\beta_v)}\mathrm{,~and} 
\label{eq-z0} \\
\bar{\tau}_z & \equiv & \frac{Z_0+R}{Z_\infty+R} \,\tau\mathrm{.}
\label{eq-omega0}
\end{eqnarray}
Equation (\ref{eq-idealz2}) describes a semi-circle in the complex
plane of radius $\frac{1}{2}\vert Z_\infty-Z_0\vert$ centered on the
real axis at $\frac{1}{2}(Z_\infty + Z_0)$.  The frequency at the peak
of the circle is given by the dynamic time constant $\bar{\tau}_z$.
This time constant is not the same as the effective time constant,
which describes the effect of electro-thermal feedback on the detector
response and is dependent on the relative resistance values of the
detector and load resistor.  The dynamic time constant is a property
of the detector only.  Equations (\ref{eq-idealz2})--(\ref{eq-omega0})
are equivalent to those given by \citet{mather84}.

When $\alpha_v > 0$ [as is the case for a superconducting
transition-edge sensor (TES)] it is possible for the impedance to
become infinite [$L_v = 1-\beta_v$ in Eq.\ (\ref{eq-z0})].  To avoid
this, one might instead measure the complex admittance $A(\omega) =
1/Z(\omega)$.  The relations describing the admittance are easily
found by transforming the preceding relations using the dual circuit
theorem, namely
\begin{eqnarray}
I & \leftrightarrow & V \label{eq-dual1} \mathrm{,}\\
R & \leftrightarrow & S \, (\, \equiv 1/R) \mathrm{,}\\
Z & \leftrightarrow & A \mathrm{,}\\
C \mathrm{~(parallel~capacitance)~} & \leftrightarrow & L \mathrm{~(series~inductance)}. \label{eq-dual2}
\end{eqnarray}
Since it is customary to keep $\alpha$ and $\beta$ as derivatives of
resistance rather than conductance, the signs of these quantities will
change:
\begin{equation}
\left. \frac{T}{S}\frac{\partial S}{\partial T}\right\vert_I = 
\left.-\frac{T}{R}\frac{\partial R}{\partial T}\right\vert_I \equiv
-\alpha_i \mathrm{,}
\end{equation}
\begin{equation}
\left.\frac{I}{S}\frac{\partial S}{\partial I}\right\vert_T =
\left.-\frac{I}{R}\frac{\partial R}{\partial I}\right\vert_T \equiv 
-\beta_i \mathrm{,}
\end{equation}
\begin{equation}
L_i = \frac{\alpha_i P}{GT} \mathrm{.}
\end{equation}
The complex admittance is then
\begin{equation}
A(\omega) = \frac{A_0+A_\infty}{2} + \frac{A_0-A_\infty}{2} \frac{1-\ij\omega\bar{\tau}_a}{1+\ij\omega\bar{\tau}_a}
\label{eq-ideala}
\end{equation}
where
\begin{eqnarray}
A_\infty & = & S/(1+\beta_i)\mathrm{,} \label{eq-ainf} \\ 
A_0 & = & A_\infty \frac{1-L_i}{1+L_i/(1+\beta_i)}\mathrm{,~and} \label{eq-a0}\\
\bar{\tau}_a & = & \frac{A_0 + S}{A_\infty+S} \,\tau\mathrm{.}
\label{eq-taua}
\end{eqnarray}

With these equations we see that the path a negative detector
($\alpha_v < 0$) traces through the complex impedance plane is
equivalent to the path traced through the complex admittance plane for
a positive detector ($\alpha_i > 0$).  


\section{Impedance Measurements} \label{sect-imped_meas}

\begin{figure*}
\includegraphics{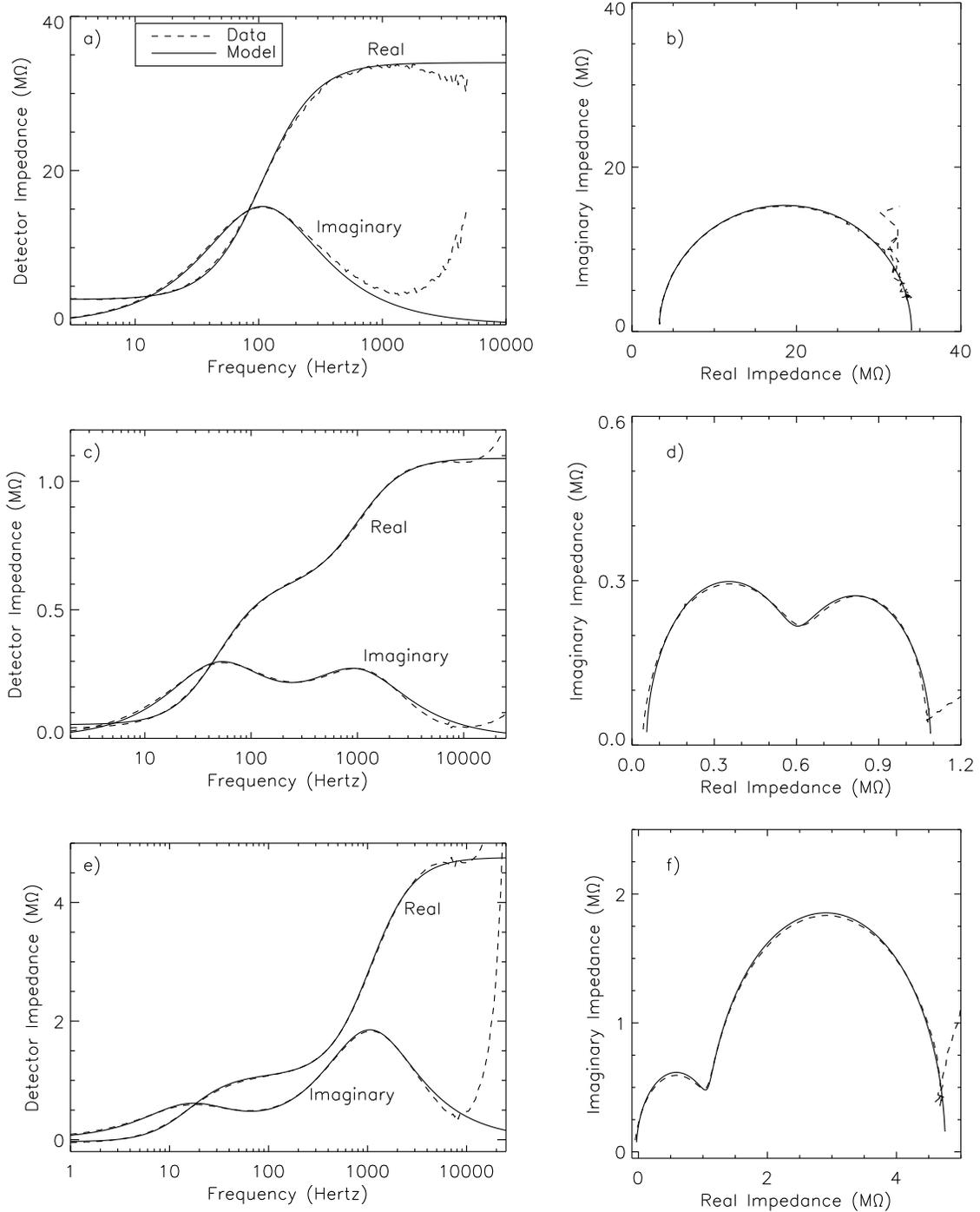}
\caption{Examples of impedance measurements for a bare thermometer
  [(a) and (b)] and thermometers with absorbers [(c)--(f)].  Data are
  shown as dotted lines and the fitted model as solid lines.  The
  figures on the left [(a), (c), and (e)] show the frequency variation
  of the real and imaginary parts of the detectors' complex impedance
  while the figures on the right [(b), (d), and (f)] plot the same
  data in the complex plane.  The divergence of the fits at high
  frequency is due to limitations of modeling the electrical circuit
  stray capacitances. The x:y axis ratio is 1:1 in (b), and 2:1 in (d)
  and (f).  For clarity, data above 5\,kHz have been removed in (a)
  and (b).
  The fit model parameters for (a)--(d) are given in Table \ref{tbl-combine}.
}
\label{fig-multiz}
\end{figure*}

Figure \ref{fig-multiz} shows examples of impedance measurements and
fitted models for three different microcalorimeter detectors.  The
first example is for a doped silicon thermistor with no absorber
[Figs.\ \ref{fig-multiz}(a) and \ref{fig-multiz}(b)].  The single
semi-circle is adequately represented by Eq.\ (\ref{eq-idealz2}) with
the parameters shown in Table \ref{tbl-combine}.  The mismatch between
the data and models at high frequency is most likely due to incomplete
modeling of stray capacitances (see Sec.\ \ref{sect-strays}).  In the
other two examples, composite absorbers have been glued to the
detectors [Figs.\ \ref{fig-multiz}(c)--\ref{fig-multiz}(f)].  The more
complex behavior of $Z(\omega)$ indicates significant thermal
resistances which can be reasonably fit by the thermal model of
Fig.\ \ref{fig:thermal_models}(b).  The complex impedance for this
model is derived by \citet{cal_model} as their Eq.\ (110) (note that
they use $\beta_i$ rather than $\beta_v$).

\begin{table}[t]
\caption{Example detector parameters
  \label{tbl-combine}}
\begin{ruledtabular}
\begin{tabular}{lccl}
  Parameter & Bare\footnote{Bare thermistor shown in Figs.\ \ref{fig-multiz}(a)--(b)} & 
  Absorber\footnote{Thermistor with absorber shown in Figs.\ \ref{fig-multiz}(c)--(d)} 
  & Method\footnote{Method used to measure the indicated parameter} \\
\hline
  $R$ (M$\Omega$) & 36 & 1.3 & dc measurement \\
  $P$ (pW) & 0.22 & 7.0 & dc measurement \\
  $T$ (mK) & 70 & 140 & $R$-vs-$T$ calibration \\
  $\alpha_v$ & -7.5 & -4.5 & $R$-vs-$T$ calibration \\
  $G$ (pW/K) & 35 & 230\footnote{Series combination of $G$ and $G_\mathrm{e}$} & dc IV-curve \\
  $C_\mathrm{p}$ (pJ/K) & \nodata & 0 & fixed \\
  $\bar{\tau}_z$ (ms) & 1.5 & \nodata & Impedance fit \\
  $Z_0$ (M$\Omega$) & 3.3 & 0.053 & Impedance fit \\
  $Z_\infty$ (M$\Omega$) & 34 & 1.1 & Impedance fit \\
  $\beta_v$  & -0.058 & -0.17 &  Eq.\ (\ref{eq-zinf}) \\
  $L_v$  & -0.83 & -0.91 & Eq.\ (\ref{eq-z0}) \\
  $\tau$ (ms) & 2.7 & \nodata & Eq.\ (\ref{eq-omega0}) \\
  $\alpha_v$ & -9.0 & -4.2 &  Eq.\ (\ref{eq-l}) \\
  $C_\mathrm{e}$ (pJ/K) & 0.093 & 0.13 & Eq.\ (\ref{eq:tau}), Impedance fit\footnote{$C_\mathrm{e}$ for the bare thermistor is found from Eq.\ (\ref{eq:tau}); it is fit directly for the thermistor with absorber.} \\
  $C_\mathrm{a}$ (pJ/K) & \nodata & 0.64 & Impedance fit \\
  $G_\mathrm{a}$ (pW/K) & \nodata & 460 & Impedance fit \\
\end{tabular}
\end{ruledtabular}
\end{table}

The complete characterization of these thermal detectors includes the
sensitivities of the thermistor $\alpha$ and $\beta$, and the heat
capacities and thermal conductivities of all constituent parts.
However, these parameters cannot be determined from measurements of
the impedance alone. The impedance of the ideal thermal model is
completely described by the three parameters $Z_0$, $Z_\infty$, and
$\bar{\tau}_z$ in Eq.\ (\ref{eq-idealz2}), which are determined by
fitting the measured $Z(\omega)$.  A dc determination of the
resistance yields $\beta_v$, $L_v$, and $\tau$ through
Eqs.\ (\ref{eq-zinf})--(\ref{eq-omega0}).  However, $\alpha_v$, $C$,
and $G$ cannot be separated from $L_v$ and $\tau$ without another
independent measurement.  For this last measurement, $G(T)$ is fitted
to dc $IV$-curves.

The data in Figs.\ \ref{fig-multiz}(c)--\ref{fig-multiz}(f) were fit
using the model in Fig.\ \ref{fig:thermal_models}(b), which separates
the detector heat capacity into contributions from the absorber
$C_\mathrm{a}$, thermistor electrons $C_\mathrm{e}$, and thermistor
phonons $C_\mathrm{p}$, connected by the labeled thermal
conductivities.  A complete characterization requires eight
parameters: three heat capacities, three thermal
conductivities, and two thermometer sensitivities.  In principle,
this model contains three separate thermal time constants which could
appear as three separate circles in the complex impedance data.  In
practice, the phonon heat capacity in our doped silicon thermistors is
much smaller than either the electron or absorber heat capacities,
placing this third time constant at frequencies beyond the practical
measurement range.  Therefore, $C_\mathrm{p} \approx 0$ and $G$ and
$G_\mathrm{e}$ can be replaced by their series combination.
The remaining six parameters are still perfectly correlated in fits to
the impedance alone so independent measurements are required to
separate them, just as in the ideal model.

The procedures outlined above for characterizing thermal detectors
contain at least one known systematic error.  The fits of $G(T)$ to
$IV$-curves assume that $\beta_v=0$, while the impedance fits and
resistance measurements clearly indicate that $\beta_v \neq0$ (see
Table \ref{tbl-combine}).  This problem could be alleviated using an
iterative procedure (re-fitting the $IV$-curves with the non-zero
$\beta_v$) or by performing simultaneous fits to both the $IV$-curves
and impedance.  Without these corrections the small values of
$\beta_v$ measured in the doped silicon thermometers
($\vert\beta_v\vert < 0.2$) introduce only a few percent uncertainty
in the $IV$-curve determination of $G(T)$.  These required corrections
may be larger in TES detectors which can have relatively large values
of $\beta_i$ ($\approx 1$--3).\citep{lindeman04}

\section{Readout Circuits} \label{sect-strays}

\subsection{The Transfer Function}

\begin{figure*}
\includegraphics[width=18cm]{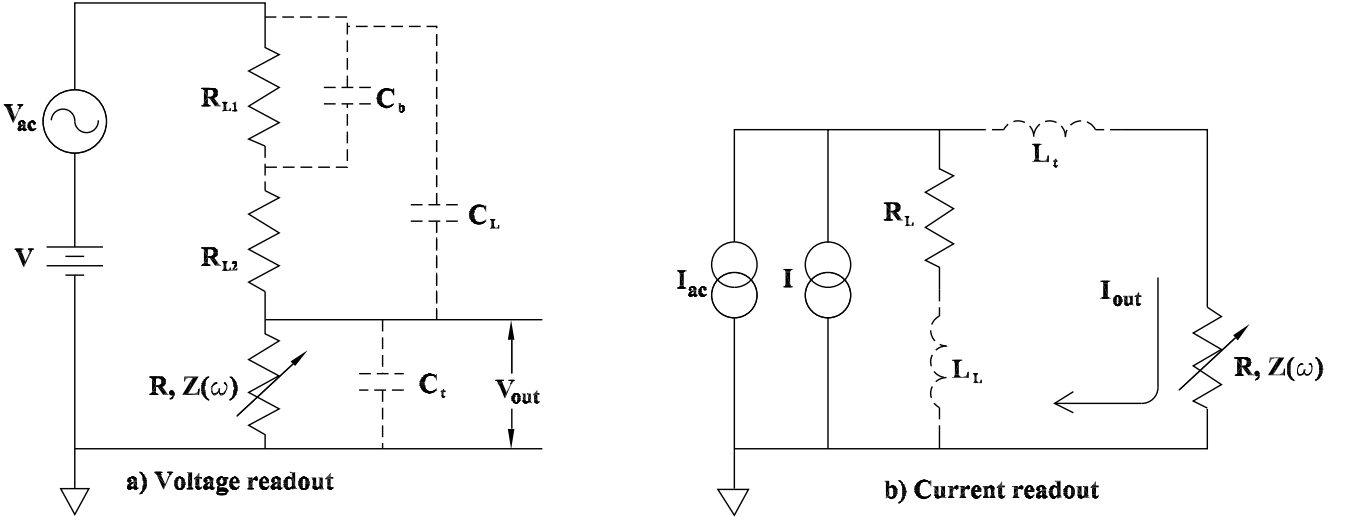}
\caption{Bias circuits for measuring detector transfer functions.
  Both circuits consist of a thermometer impedance $Z(\omega)$,
  resistance $R$, and load resistor $R_\mathrm{L}$. a) Voltage
  readout circuit for doped silicon thermometers.  The capacitors
  connected by dotted lines are used to model stray capacitances in
  the system.  $C_\mathrm{L}$ and $C_\mathrm{t}$ are shunt
  capacitances across the load resistor and thermometer, respectively.
  The load resistor is one physical resistor, but it has been split in
  this schematic ($R_\mathrm{L} = R_{L_1} +R_{L_2}$) in order to
  partially model the distributed capacitances $C_\mathrm{b}$.  b)
  Current readout circuit for TES\@.  The inductors shown as dotted
  lines are used to model stray inductances in series with the
  thermometer, $L_\mathrm{t}$, and load resistor, $L_\mathrm{L}$.  If
  $C_\mathrm{b} = 0$ these two circuits are dual equivalents and all
  equations describing one system can be transformed to the other
  system using the substitutions in Eqs.\
  (\ref{eq-dual1})--(\ref{eq-dual2}).
\label{fig-circuits}}
\end{figure*}

Figure \ref{fig-circuits} illustrates two equivalent circuits for
biasing thermal detectors, either of which can be used to measure the
impedance.  The first utilizes voltage readout
[Fig.\ \ref{fig-circuits}(a)], the other current readout
[Fig.\ \ref{fig-circuits}(b)].
Our technique for measuring the dynamic impedance is to add a small
ac signal to the dc bias and measure the resulting complex
transfer function as a function of frequency.  The transfer functions
are defined as $T(\omega) \equiv
V_\mathrm{out}(\omega)/V_\mathrm{in}(\omega)$ for voltage readout and
$T(\omega) \equiv I_\mathrm{out}(\omega)/I_\mathrm{in}(\omega)$ for
current readout.  The ac signal can be either a random noise source
with a bandwidth spanning the frequencies of interest or a sinusoidal
source that can be scanned through the desired frequency range.  We
use a commercial spectrum analyzer to measure the complex transfer
function using both of these methods.  The sine-sweep method could
also be implemented with a two-phase lock-in amplifier and swept sine
generator.

\subsection{Electrical Strays} \label{sect-straycs}

For an ideal bias circuit the transfer function would simply be the
voltage or current divider formed by $R_L$ and $Z(\omega)$.  In
practice, additional reactive components exist in the form of stray
capacitances and/or inductances.  These stray reactances will
introduce additional phase and amplitude shifts in $T(\omega)$, but it
is possible to correct for these effects if the strays in the bias
circuit can be accurately modeled and measured.

The stray reactances can be determined by measuring the transfer
function of the circuit with the reactive part of $Z(\omega)$ removed.
This is done by either replacing the detector with a pure resistor or
by measuring the transfer function of an unbiased ($\langle
V_\mathrm{in}\rangle = \langle I_\mathrm{in}\rangle = 0$) detector.
If the unbiased transfer function is measured with a sufficiently
small ac signal, such that the detector dissipates negligible power
($P \approx 0$), then the detector is purely resistive [see
Eqs.\ (\ref{eq-idealz1}) and (\ref{eq-l})].  Any observed reactive
component under these conditions must be due to stray reactances in
the circuit.

The dominant reactances in our voltage readout circuit for silicon
detectors are shunt capacitances across the detector and load resistor
($C_\mathrm{t}$ and $C_\mathrm{L}$ in Fig.\ \ref{fig-circuits},
respectively).  For this model circuit the transfer function is given
by
\begin{equation}
T(\omega) = \frac{V_\mathrm{out}}{V_\mathrm{in}} = \frac{Z(\omega)}{R_\mathrm{L} + Z(\omega)} \, \frac{1 + \ij\omega\tau_\mathrm{L}}{1 + \ij\omega\tau_\mathrm{t}}
\label{eq-vtf}
\end{equation}
where
\begin{eqnarray}
\tau_\mathrm{L} & = & R_\mathrm{L} C_\mathrm{L} \qquad \mathrm{and} \\
\tau_\mathrm{t} & = & \frac{C_\mathrm{L} + C_\mathrm{t}}{R_\mathrm{L}^{-1} + Z(\omega)^{-1}}.
\label{eq-tftau}
\end{eqnarray}
This simple model for the electrical strays also traces a semi-circle
in the complex plane if the detector impedance is completely real,
$Z(\omega) = R$.
\begin{equation}
T(\omega) = \frac{T_0+T_\infty}{2} + \frac{T_0-T_\infty}{2} \frac{1-\ij\omega\tau_\mathrm{t}}{1+\ij\omega\tau_\mathrm{t}}
\label{eq-tf}
\end{equation}
where
\begin{eqnarray}
T_\infty & \equiv & \lim_{\omega\rightarrow\infty} T(\omega) =
\frac{C_\mathrm{L}}{C_\mathrm{L} + C_\mathrm{t}}\mathrm{,~and}
\label{eq-tinf} \\
T_0 &\equiv & T(0) = \frac{R}{R_\mathrm{L} + R} 
\label{eq-t0}
\end{eqnarray}

The dual circuit theorem can be used to transform these equations from
the voltage readout circuit of Fig.\ \ref{fig-circuits}(a) to the
current readout circuit of Fig.\ \ref{fig-circuits}(b).  Equations
(\ref{eq-vtf})--(\ref{eq-tftau}) are then
\begin{eqnarray}
T(\omega) = \frac{I_\mathrm{out}}{I_\mathrm{in}} & = &
\frac{A(\omega)}{S_\mathrm{L} + A(\omega)} \, \frac{1 +
\ij\omega\tau_\mathrm{L}}{1 + \ij\omega\tau_\mathrm{t}} \nonumber \\
& = & 
\frac{R_\mathrm{L}}{R_\mathrm{L} + Z(\omega)} \, \frac{1 +
\ij\omega\tau_\mathrm{L}}{1 + \ij\omega\tau_\mathrm{t}}
\end{eqnarray}
where
\begin{eqnarray}
\tau_\mathrm{L} & = & L_\mathrm{L} / R_\mathrm{L} \qquad \mathrm{and}
\\ \tau_\mathrm{t} & = & \frac{L_\mathrm{L} + L_\mathrm{t}}{R_\mathrm{L} + Z(\omega)}.
\label{eq-itf}
\end{eqnarray}
Equations (\ref{eq-tf})--(\ref{eq-t0}) can be similarly transformed.

\begin{figure}[!t]
\includegraphics[width=9cm]{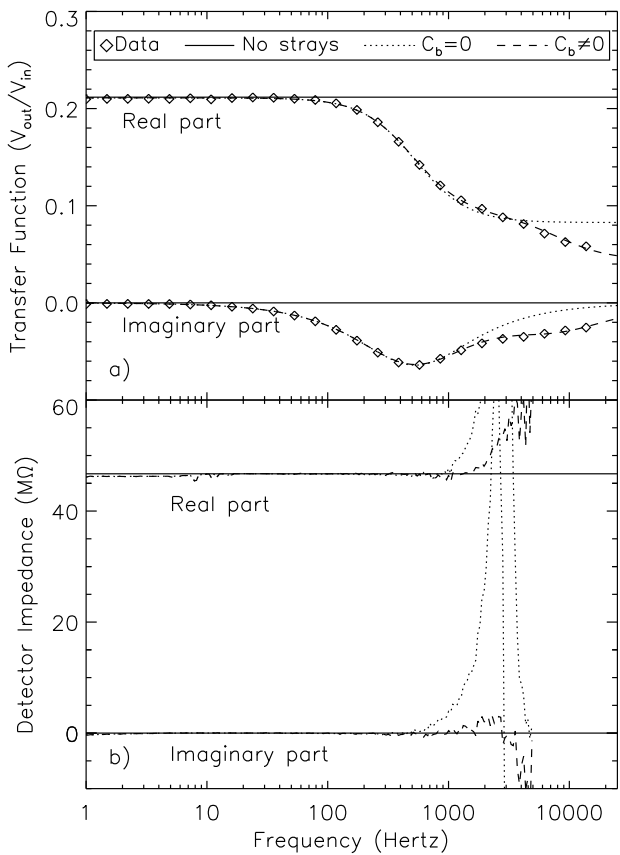}
\caption{Example of an unbiased transfer function.  a) The measured
  complex transfer function (diamonds represent every tenth data
  point) and fits to two stray capacitance models [dotted and dashed
    lines; Fig.\ \ref{fig-circuits}(a)].  If no strays were present
  the data would follow the solid lines.  b) The inferred detector
  impedance after correcting for each stray model (dotted and dashed
  lines) is compared with an ideal resistor (solid lines).  One model
  uses only shunt capacitances across the detector and load resistor
  ($C_\mathrm{b} = 0$) while the other also includes a distributed
  capacitance within the load resistor ($C_\mathrm{b} \neq 0$).  The
  impedance data above 5\,kHz have been removed from the bottom plot
  for clarity.
\label{fig-unbias}}
\end{figure}

Real strays are complex, involving distributed reactances rather than
(or in conjunction with) the simple parallel capacitances and series
inductances discussed here.  It is the accuracy of the circuit model,
not the values of the circuit elements (both real and stray) that
limits the maximum useful frequency for impedance measurements.  This
will particularly limit the accuracy with which $Z_\infty$ and
$\beta_v$ (or $A_\infty$ and $\beta_i$) can be measured.

No stray inductances have been included in Fig.\ \ref{fig-circuits}(a)
as they are generally negligible for high-impedance silicon detectors.
Similarly, the effect of stray capacitances can be neglected for the
very low impedance TES detectors.  In this case the stray inductances
shown in Fig.\ \ref{fig-circuits}(b) can be significant.  However,
these stray inductances still seem to be less of a problem for the
low-impedance detectors than stray capacitances are for high-impedance
detectors.  As a result, the impedance of a TES can often be measured
up to higher frequencies (tens of kHz)\citep{lindeman04,saabltd10}
than the silicon detectors (a few kHz).

We note that the utility of the impedance method is not limited to
only those systems which can be described by the bias circuit models
in Figure \ref{fig-circuits}.  It is easily extended to any bias circuit
whose frequency dependent transfer function can be modeled and
measured. This includes bridge circuits and long transmission lines,
such as those often employed in fusion
research.\citep{paul93,schivell82,murari04,giannone02}

\begin{figure}[!t]
\includegraphics[width=9cm]{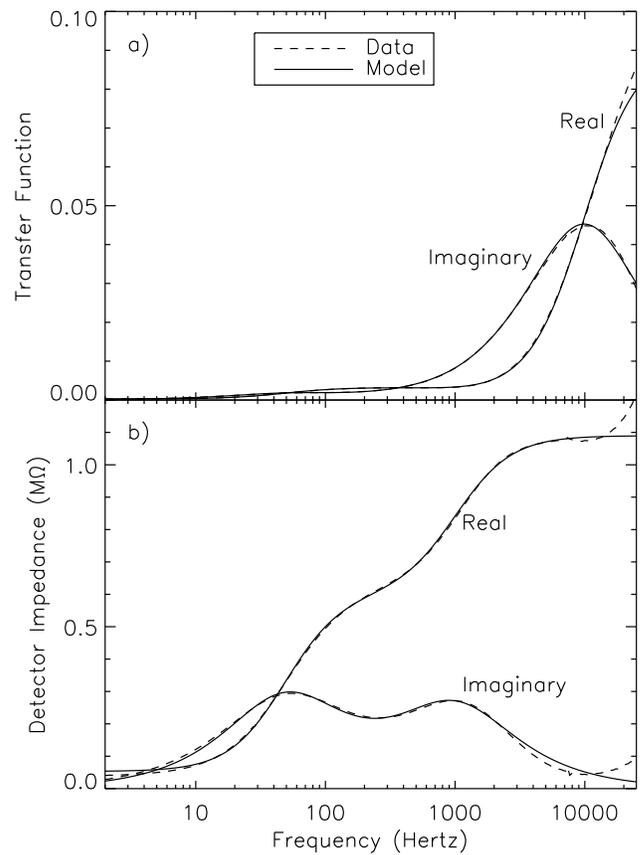}
\caption{Measured and fitted transfer functions of a biased detector
with absorber. a) Transfer function. b) Detector impedance inferred
from the transfer function using the measured stray capacitances.
\label{fig-absorberz}}
\end{figure}

\subsection{Measuring Stray Capacitances}

The shunt capacitance across the silicon detectors is typically
$C_\mathrm{t} \sim 10$\,pF\@.  The load resistors used in this work
are made of nickel-chromium thin film deposited on a thin silicon
nitride insulating layer on a degenerate silicon
substrate\footnote{purchased from Mini Systems, Inc., Attleboro, MA
  02703}. These were found to have shunt capacitances on the order of
$C_\mathrm{L} \sim 1$\,pF\@.  A set of Nichrome resistors on quartz
substrates\addtocounter{footnote}{-1}\footnotemark\ had shunt
capacitances an order of magnitude smaller.

These two shunt capacitances alone are not always sufficient to fit
the unbiased transfer functions at high frequencies.  In an effort to
better model the distributed capacitance in the load resistor the
resistance is split into two pieces and a third shunt capacitor is
added, $C_\mathrm{b}$ [Fig.\ \ref{fig-circuits}(a)].
Typical values of these additional components are $C_b \lesssim 1$\,pF
and $R_{L_1}/R_{L_2} \approx 1$ -- 5.

Example fits of an unbiased transfer function are shown in Fig.\
\ref{fig-unbias}.  Including the effects of the stray capacitances in
fitting the unbiased data should result in a completely real
(resistive) impedance which is independent of frequency.  The lower
panel of Fig.\ \ref{fig-unbias} makes it clear that our simplest
stray model ($C_\mathrm{b} = 0$) accomplishes this only at frequencies
below 500\,Hz.  If we include the distributed capacitance in the load
resistor ($C_\mathrm{b} \neq 0$) the effects can be removed up to
frequencies of $\approx 1$\,kHz.  The remaining deviations above 1\,kHz
indicate that our stray models are not an adequate representation of
the real circuit at these frequencies.


\subsection{Biased Transfer Functions}

In Fig.\ \ref{fig-absorberz}(a) we measure the transfer function of a
biased detector (dotted lines).  Using this transfer function and the
measured values of the load resistor and stray capacitances we can
calculate the impedance of the thermal detector from
Eq.\ (\ref{eq-vtf}) or its equivalent; this impedance is shown as the
dotted lines in Fig.\ \ref{fig-absorberz}(b).  The best fit to the
biased transfer function [solid line in Fig.\ \ref{fig-absorberz}(a)]
is obtained by varying the detector parameters (i.e.\ $\alpha$,
$\beta$, $C$'s, $G$'s) while the strays remain fixed at their measured
values.  The impedance for these best fit detector parameters is shown
as the solid lines in Fig.\ \ref{fig-absorberz}(b). As already
mentioned in Sec.\ \ref{sect-imped_meas}, the divergence at high
frequencies is most likely due to inadequate modeling of the stray
capacitances.

%


\begin{acknowledgments}
The doped silicon thermometers and their absorbers were fabricated by
Regis Brekosky, Caroline Kilbourne, and colleagues at NASA/Goddard
Space Flight Center.  We thank Lindsay Rocks, Dahai Liu, and Melanie
Clarke for their contributions to data analysis and acquisition.  We
would also like to thank Wilt Sanders, Enectali Figueroa-Feliciano,
and Massimiliano Galeazzi for many useful discussions and especially
Dan McCammon and Mark Lindeman for discussions and careful readings of
this manuscript.  This work has been supported by NASA grant
NAG5-5404.
\end{acknowledgments}


\end{document}